\newenvironment{keywords}{\centerline{\bf\small
Keywords}\begin{quote}\small}{\par\end{quote}\vskip 1ex}

\documentclass[12pt,twoside]{article}
\begin{document}

\title{\vspace{-4ex}
\vskip 2mm\bf\Large\hrule height5pt \vskip 4mm
Algorithmic Randomness as Foundation of Inductive Reasoning and Artificial Intelligence
\vskip 4mm \hrule height2pt}
\author{{\bf Marcus Hutter}\\[3mm]
\normalsize SoCS, RSISE, IAS, CECS \\[-0.5ex]
\normalsize Australian National University \\[-0.5ex]
\normalsize Canberra, ACT, 0200, Australia \\
}
\date{5 May 2010}
\maketitle

\begin{abstract}
This article is a brief personal account of the past, present, and future
of algorithmic randomness, emphasizing its role in inductive
inference and artificial intelligence. It is written for a general
audience interested in science and philosophy.
Intuitively, randomness is a lack of order or predictability. If
randomness is the opposite of determinism, then algorithmic
randomness is the opposite of computability. Besides many other
things, these concepts have been used to quantify Ockham's razor,
solve the induction problem, and define intelligence.
\def\contentsname{\centering\normalsize Contents}
{\parskip=-2.7ex\tableofcontents}
\end{abstract}

\begin{keywords}\vspace*{-1ex}
algorithmic information theory;
individual randomness;
Ockham's razor;
inductive reasoning;
artificial intelligence.
\end{keywords}

\newpage
\section{Why were you initially drawn to the study of computation and randomness?} 

Some sequences of events follow a long causal ``computable'' path,
while others are so ``random'' that the coherent causal path is
quite short. I am able to trace back quite far my personal causal
chain of events that eventually led me to computability and
randomness (C\&R), although the path looks warped and random.

At about 8 years of age, I got increasingly annoyed at always having to tidy
up my room. It took me more than 20 years to see that
computation and randomness was the solution to my problem
(well -- sort of). Here's a summary of the relevant events:

First, my science fiction education came in handy. I was well-aware that
robots were perfectly suitable for all kinds of boring jobs, so they
should be good for tidying up my room too. Within a couple of years
I had built a series of five increasingly sophisticated robots. The
``5th generation'' one was humanoid-shaped, about 40cm high, had two
arms and hands, and one broad roller leg. The frame was metal and
the guts cannibalized from my remote controlled car and other toys.

With enough patience I was able to maneuver Robbie5 with the remote
control to the messy regions of my room, have it pick up some Lego pieces
and place them in the box they belonged to. It worked! And it was
lots of fun. But it didn't really solve my problem. Picking up a block
with the robot took at least 10 times longer than doing it by hand,
and even if the throughput was the same, I felt I hadn't gained much.

Robbie5 was born abound a year before my father brought home one of
the first programmable pocket calculators in 1978, a HP-19C. With
its 200 bytes of RAM or so it was not quite on par with Colossus (a
super computer which develops a mind of its own in the homonymous
movie), but HP-19C was enough for me to realize that a computer
allows programming of a robot to perform a sequence of steps
autonomously.
Over the following 15 years, I went through a sequence of
calculators and computers, wrote increasingly sophisticated
software, and studied computer science with a Masters degree in
Artificial Intelligence (AI). My motivation in AI of course changed
many times over the years, from the dream of a robot tidying up my
room to more intellectual, philosophical, economic, and social
motivations.

Around 1992 I lost confidence in any of the existing approaches
towards AI, and despite considerable effort for years, didn't have
a ground-breaking idea myself.

While working in a start-up company on a difficult image
interpolation problem, I realized one day in 1997
that simplicity and compression are key, not only for solving my
problem at hand, but also for the grand AI problem.

It took me quite a number of weekends to work out the details.
Relatively soon I concluded that the theory I had developed was too
beautiful to be novel. I had rediscovered aspects of Kolmogorov
complexity and Solomonoff induction. Indeed, I had done more. My
system generalized Solomonoff induction to a universally optimal
general-purpose reinforcement learning agent.

In order to prove some of my claims it was necessary to become more
deeply and broadly acquainted with Algorithmic Information Theory
(AIT).

AIT combines information theory and computation theory to an
objective and absolute notion of information in an individual
object, and gives rise to an objective and robust notion of
randomness of individual objects. Its major sub-disciplines are
Algorithmic ``Kolmogorov'' Complexity (AC), Algorithmic
``Solomonoff'' Probability (AP), Algorithmic ``Martin-L\"of''
Randomness (AR), and Universal ``Levin'' Search (UL)
\cite{Hutter:07ait}.

This concludes my 25(!)\ year journey to C\&R. In the last 10 years,
I have contributed to all the 4 subfields. My primary driving force
when doing research in C\&R is still AI, so I've most to say about
AP, and my answers to the following questions are biased towards my
own personal interests.

\section{What have we learned?} 

Let me begin with what {\em I} have learned: The most important
scientific insight I have had is the following: Many scientists have a
bias towards elegant or beautiful theories, which usually aligns
with some abstract notion of simplicity. Others have a bias towards
simple theories in the concrete sense of being analytically or
computationally tractable. By `theories' I essentially mean
mathematical models of some aspect of the real world, e.g.\ of a
physical system like an engine or the weather or stock market.

Way too late in my life, at age 30 or so, I realized that the
most important reason for preferring simple theories is a quite
different one: Simple theories tend to be better for what they are
developed for in the first place, namely predicting in related but
different situations and using these predictions to improve
decisions and actions.

Indeed, the principle to prefer simpler theories has been popularized
by William of Ockham (1285-1349) (``Entities should not be
multiplied unnecessarily'') but dates back at least to Aristotle
\cite{Franklin:02}.

Kolmogorov complexity \cite{Kolmogorov:65} is a universal objective
measure of complexity and allows simplicity and hence Ockham's
``razor'' principle to be quantified. Solomonoff
\cite{Solomonoff:64} developed a formal theory of universal
prediction along this line, actually a few years before Kolmogorov
introduced his closely related complexity measure. My contribution
in the 200X \cite{Hutter:00kcunai,Hutter:07aixigentle} was to
generalize Solomonoff induction to a universally intelligent
learning agent \cite{Oates:06}.

This shows that Ockham's razor, inductive reasoning, intelligence,
and the scientific inquiry itself are intimately related. I would
even go so far as to say that science {\em is} the application of
Ockham's razor: {\em Simple} explanations of observed real-world
phenomena have a higher chance of leading to correct predictions
\cite{Hutter:09ctoex}.

What does all this have to do with C\&R? We cannot be certain about
{\em anything} in our world. It might even end or be completely
different tomorrow. Even if some proclaimed omniscient entity told
us the future, there is no scientific way to verify the premise of
its omniscience. So induction has to deal with uncertainty. Making
worst-case assumptions is not a generic solution; the generic
worst-case is ``anything can happen''. Considering restricted model
classes begs the question about the validity of the model class
itself, so is also not a solution. More powerful is to model
uncertainty by probabilities and the latter is obviously related to
randomness.

There have been many attempts to formalize probability and
randomness: Kolmogorov's axioms of probability theory
\cite{Kolmogorov:33} are the default characterization. Problems with
this notion are discussed in item (d) of Question 4. Early attempts
to define the notion of randomness of {\em individual}
objects/sequences by von Mises \cite{VonMises:19}, Wald
\cite{Wald:37}, and Church \cite{Church:40} failed, but finally
Martin-L\"of \cite{MartinLoef:66} succeeded. A sequence is
Martin-L\"of random if and only if it passes all effective
randomness tests or, as it turns out, if and only if it is
incompressible.

\section{What don't we know (yet)?} 

Lots of things, so I will restrict myself to open problems in the
intersection of AIT and AI. See \cite{Hutter:09aixiopen} for details.

(i) Universal Induction: The induction problem is a fundamental
problem in philosophy \cite{Hume:1739,Earman:93} and statistics
\cite{Jaynes:03} and science in general. The most important
fundamental philosophical and statistical problems around induction
are discussed in \cite{Hutter:07uspx}: Among others, they include
the problems of old evidence, ad-hoc hypotheses, updating, zero
prior, and invariance. The arguments in \cite{Hutter:07uspx} that
Solomonoff's universal theory $M$ overcomes these problems are
forceful but need to be elaborated on further to convince the
(scientific) world that the induction problem is essentially solved.
Besides these general induction problems, universal induction raises
many additional questions: for instance, it is unclear whether $M$
can predict all computable {\em sub}sequences of a sequence that is
itself not computable, how to formally identify ``natural'' Turing
machines \cite{Mueller:10}, Martin-L\"of convergence of $M$, and
whether AIXI (see below) reduces to $M$ for prediction.

(ii) Universal Artificial Intelligence (UAI): The AIXI model
integrates Solomonoff induction with sequential decision theory. As
a unification of two optimal theories in their own domains, it is
plausible that AIXI is optimal in the ``union'' of their domains.
This has been affirmed by positive pareto-optimality and
self-optimizingness results \cite{Hutter:04uaibook}. These results
support the claim that AIXI is a universally optimal generic
reinforcement learning agent, but unlike the induction case, the
results so far are not yet strong enough to allay all doubts.
Indeed, the major problem is not to {\em prove} optimality but to
{\em come up} with sufficiently strong but still satisfiable
optimality notions in the reinforcement learning case.
A more modest goal than proving optimality of AIXI is to ask for
additional reasonable convergence properties, like posterior
convergence for unbounded horizon.
The probably most important fundamental and hardest problem in game
theory is the grain-of-truth problem \cite{Kalai:93}. In our
context, the question is what happens if AIXI is used in a
multi-agent setup \cite{Shoham:08} interacting with other
instantiations of AIXI.

(iii) Defining Intelligence:
A fundamental and long standing difficultly in the field of AI is
that intelligence itself is not well defined.
Usually, formalizing and rigorously defining a previously vague
concept constitutes a quantum leap forward in the corresponding
field, and AI should be no exception.
AIT again suggested an extremely general, objective, fundamental,
and formal measure of machine intelligence
\cite{Hutter:00kcunai,Legg:08,Graham-Rowe:05,Fievet:05},
but the theory surrounding it has yet to be adequately explored.
A comprehensive collection, discussion and comparison of
verbal and formal intelligence tests, definitions, and measures
can be found in \cite{Hutter:07iorx}.

\section{What are the most important open problems in the field?} 

There are many important open technical problems in AIT. I have
discussed some of those that are related to AI in
\cite{Hutter:09aixiopen} and in the previous answer. Here I
concentrate on the most important open problems in C\&R which I am
able to describe in non-technical terms.

(a) The development of notions of complexity and individual randomness
didn't end with Kolmogorov and Martin-L\"of.
Many variants of ``plain'' Kolmogorov complexity $C$ \cite{Kolmogorov:65} have been developed:
prefix complexity $K$ \cite{Levin:74,Gacs:74,Chaitin:75}, %
process complexity \cite{Schnorr:73}, %
monotone complexity $K\!m$ \cite{Levin:73random}, %
uniform complexity \cite{Loveland:69ic,Loveland:69acm}, %
Chaitin complexity $K\!c$ \cite{Chaitin:75}, %
Solomonoff's universal prior $M=2^{-K\!M}$ \cite{Solomonoff:64,Solomonoff:78}, %
extension semimeasure $M\!c$ \cite{Cover:74}, %
and some others \cite{Li:08}. %
They often differ only by $O(\log K)$,
but this can lead to important differences.
Variants of Martin-L\"of randomness are:
Schnorr randomness \cite{Schnorr:71},
Kurtz randomness \cite{Kurtz:81},
Kolmogorov-Loveland randomness \cite{Loveland:66}, %
and others \cite{Wang:96,Calude:02,Downey:07book}.
All these complexity and randomness classes can further be relativized to
some oracle, e.g.\ the halting oracle, leading to an arithmetic
hierarchy of classes.
Invoking resource-bounds moves in the other direction and leads to
the well-known complexity zoo \cite{Aaronson:05} and
pseudo-randomness \cite{Luby:96}.
%
Which definition is the ``right'' or ``best'' one, and in which
sense?
Current research on algorithmic randomness is more concerned about
abstract properties and convenience, rather than practical usefulness. This is
in marked contrast to complexity theory, in which the classes also
sprout like mushrooms \cite{Aaronson:05}, but the majority of
classes delineate important practical problems.

(b) The historically oldest, non-flawed, most popular, and default
notion of individual randomness is that of Martin-L\"of. Let us
assume that it is or turns out to be the ``best'' or single
``right'' definition of randomness. This would uniquely determine
which individual infinite sequences are random and which are not.
This unfortunately does not hold for finite sequences. This
non-uniqueness problem is equivalent to the problem that Kolmogorov
complexity depends on the choice of universal Turing machine. While
the choice is asymptotically, and hence for large-scale practical
applications, irrelevant, it seriously hinders applications to
``small'' problems. One can argue the problem away
\cite{Hutter:04uaibook}, but finding a unique ``best'' universal
Martin-L\"of test or universal Turing machine would be more
satisfactory and convincing. Besides other things, it would make
inductive inference absolutely objective.

(c) Maybe randomness can, in principle, only be relative: What looks
random to me might be order to you. So randomness depends on the
power of the ``observer''. In this case, it is important to study
problem-specific randomness notions, and clearly describe and
separate the application domains of the different randomness
notions,
like classical sufficient statistics depends on the model class.
Algorithmic randomness usually includes all computable tests and
goes up the arithmetic hierarchy. For practical applications,
limited classes, like all efficiently computable tests, are more
relevant. This is the important domain of pseudo-random number
generation. Could every practically useful complexity class
correspond to a randomness class with practically relevant
properties?

(d) It is also unclear whether algorithmic randomness or classical
probability theory has a more fundamental status.
While measure theory is mathematically, and statistics is
practically very successful, Kolmogorov's probability axioms are
philosophically crippled and, strictly speaking, induce a purely
formal but meaningless measure theory exercise. The easygoing
frequentist interpretation is circular: The probability of head is
$p$, if the long-run relative frequency tends to $p$ almost surely
(with probability one). But what does `almost surely' mean?
Applied statistics implicitly invokes Cournot's somewhat forgotten
principle: An event with very small probability, singled out in
advance, will not happen. That is, a probability 1 event will
happen for sure in the real world.
Another problem is that it is not even possible to ask the question
of whether a particular single sequence of observations is random
(w.r.t.\ some measure). Algorithmic randomness makes this question
meaningful and answerable. A downside of algorithmic randomness
is that not every set of measure 1 will do, but only constructive
ones, which can be much harder to find and sometimes do not exist
\cite{Hutter:07mlconvxx}.

(e) Finally, to complete the circle, let's return to my original
motivation for entering this field: Ockham's razor (1) is the key
philosophical ingredient for solving the induction problem and
crucial in defining science and intelligence, and (2) can be
quantified in terms of algorithmic complexity which itself is
closely related to algorithmic randomness.
The formal theory of universal induction
\cite{Solomonoff:78,Hutter:07uspx} is already well-developed and the
foundations of universal AI have been laid \cite{Hutter:04uaibook}.
Besides solving specific problems like (i)-(iii) and (a)-(d) above,
it is also important to ``translate'' the results and make them
accessible to researchers in other disciplines: present the
philosophical insights in a less-mathematical way; stress that sound
mathematical foundations are crucial for advances in most field, and
induction and AI should be no exception; etc.

\section{What are the prospects for progress?} 

The prospects for the open problems (a)-(e) of Question 4 I believe
are as follows:

(a) I am not sure about the fate of the multitude of different
randomness notions. I can't see any practical relevance for those in
the arithmetic hierarchy. Possibly the acquired scientific knowledge
from studying the different classes and their relationship can be
used in a different field in an unexpected way. For instance, the
ones in the arithmetic hierarchy may be useful in the endeavor of
unifying probability and logic \cite{Gaifman:82}. Possibly the whole
idea of {\em objectively} identifying individually which strings
shall be regarded as random will be given up.

(b) All scientists, except some logicians studying logic,
essentially use the same classical logic and axioms, namely ZFC,
to do {\em deductive} reasoning. Why do not all scientists
use the same definition of probability to do {\em inductive}
reasoning?
Bayesian statistics and Martin-L\"of randomness are the most
promising candidates for becoming universally accepted for inductive
reasoning.

Maybe they will become universally accepted some time in the future,
for pragmatic reasons, or simply as a generally agreed upon
convention, since no one is interested in arguing over it anymore.
While Martin-L\"of uniquely determines infinite random sequences,
the randomness for finite sequences depends on the choice of
universal Turing machine. Finding a unique ``best'' one (if
possible) is, in my opinion, the most important open problem in
algorithmic randomness. A conceptual breakthrough would be needed to
make progress on this hard front. See \cite{Mueller:10} for a
remarkable but failed recent attempt.

(c) Maybe pursuing a single definition of randomness is illusory.
Noise might simply be that aspect of the data that is not useful for
the particular task or method at hand. For instance, sufficient
statistics and pseudo-random numbers have this task-dependence. Even
with a single fundamental notion of randomness (see b) there will be
many different practical approximations. I expect steady progress on
this front.

(d) Bayesian statistics based on classical probability theory is
incomplete, since it does not tell you how to choose the prior.
Solomonoff fixes the prior to a negative exponential in the model
complexity. Time and further research will convince classical
statisticians to accept this (for them now) exotic choice as a kind
of ``gold standard'' (as Solomonoff put it). All this is still
within the classical measure theoretic framework, which may be
combined with Cournot {\em or} with Martin-L\"of.

(e) Finally, convincing AI researchers and philosophers about the
importance of Ockham's razor, that algorithmic complexity is a
suitable quantification, and that this led to a formal (albeit
non-computable) conceptual solution to the induction and the AI problem should
be a matter of a decade or so.


\begin{small}

\end{small}

\end{document}